\documentclass[aps,prd,twocolumn,showpacs,groupedaddress]{revtex4-1}
 
\pdfoutput=1
\usepackage{amssymb}
\usepackage{amsmath}
\usepackage{hyperref}
\usepackage{graphics}

\begin{document}

\title{Energy extremum principle for charged black holes}

 \author{Scott Fraser}
\email[Email address: ]{scfraser@calpoly.edu}
\affiliation{Department of Physics, 
California Polytechnic State University,  
San Luis Obispo, California 93407, USA}

\author{Shaker Von Price Funkhouser}
\email[Email address: ]{sfunkhou@calpoly.edu}
\affiliation{Department of Physics, 
California Polytechnic State University,  
San Luis Obispo, California 93407, USA}

\begin{abstract}
For a set of $N$ asymptotically flat black holes with arbitrary charges and masses, all initially at rest and well separated, we prove the following extremum principle: the extremal charge configuration ($|q_i|=m_i$ for each black hole) can be derived by extremizing the total energy, for variations of the black hole apparent horizon areas, at fixed charges and fixed Euclidean separations.  We prove this result through second order in an expansion in the inverse separations.  If all charges have the same sign, this result is a variational principle that reinterprets the static equilibrium of the Majumdar-Papapetrou-Hartle-Hawking solution as an extremum of total energy, rather than as a balance of forces; this result augments a list of related variational principles for other static black holes, and is consistent with the independently known Bogomol'nyi-Prasad-Sommerfield (BPS) energy minimum.
    \end{abstract}

\pacs{04.70.Bw, 04.40.Nr}

\maketitle

\section{INTRODUCTION \label{introduction}}

A system  of multiple  black holes  is  not   static in general, but  
the system can be  kept static      by  a strut  (a conical singularity) on the axis between each pair of black holes.  Each   strut  is interpreted as  providing a force that prevents the black holes from   moving
 (see \cite{equilibrium-simple} and references therein). 
Remarkably, 
without any struts present, there is   also a well-known solution   \cite{Hartle-Hawking}  that  describes 
a  set of $N$   charged black holes   in static equilibrium,
regardless of their  mutual separations.  A key feature of this static solution is
  that the black holes' individual charges  $q_i$  all  have the same sign, and are related to their individual masses $m_i$ by  the  
condition (in   units with $G=c=4\pi\epsilon_0=1$)
 \begin{equation}
 \label{static condition intro}
|q_i|=m_i , \qquad 1 \le i \le N.
\end{equation}
    The asymptotically flat  static geometry satisfying   (\ref{static condition intro})
    was found        by
  Hartle and Hawking
  \cite{Hartle-Hawking}, who extended  the    Majumdar-Papapetrou  solutions   \cite{Majumdar,Papapetrou}. 
The interpretation and significance of the condition  (\ref{static condition intro})  
is of central interest in   this paper.
The traditional interpretation of the  
  Hartle-Hawking  solution   \cite{Hartle-Hawking}  appeals to forces: the static equilibrium 
   is attributed to the  exact balance between the black holes'  pairwise gravitational attraction and   electric repulsion.  
  However,  the condition  (\ref{static condition intro}) refers to each black hole, not to a balance of forces between black holes.  Thus, (\ref{static condition intro})  is   a stronger condition than  a Newtonian balance of forces, which for two particles    requires  $q_1q_2 = m_1m_2$.  This suggests that   (\ref{static condition intro}) can be interpreted without using forces.

Such an alternative interpretation is the purpose of this paper. We   show that  the   condition (\ref{static condition intro})  can be derived from  an  extremum of  the system's total energy $E$.  
 This         fits   naturally into the framework of  general relativity, where gravity and   total energy are determined by spacetime geometry, and   gravity  is fundamentally  not  treated as   a force.

The condition  (\ref{static condition intro})  describes   \textit{extremal}  charge, since a    nonrotating black hole   satisfies a charge-mass inequality,  $|q| \le m$. Thus,
 our results can be  briefly summarized   as:   
a set of extremally charged black holes extremizes  the  total energy. 
More precisely,  we consider a conformally flat spatial geometry in the form introduced by  Brill and Lindquist \cite{Brill-Lindquist}, and we prove the following extremum principle for well-separated black holes:

\begin{equation}
\parbox[b]{2.85in}{
\textit
 {For   $N$    charged  black holes, initially at rest, each with $|q_i| \le m_i$ and apparent horizon area $A_i$,  the 
 extremal charge   
 condition $|q_i| = m_i$  
follows from  extremizing  the total energy: $\partial E/\partial A_i=0$ at fixed       charges $q_i$    and   Euclidean separations  $r_{ij}$.
}
}
\label{extremum principle}
\end{equation}
We    prove the extremum principle (\ref{extremum principle}) as an expansion in the inverse   separation distances ($1/r_{ij}$), 
through second order,  which    
   is   where relativistic post-Newtonian   contributions first appear. 
The particular quantities  that are   varied or  held  fixed  in this energy extremum  are motivated by the first law of black hole mechanics, as we will illustrate when we  prove   (\ref{extremum principle}).

If all    of the black hole charges in (\ref{extremum principle})  have the same sign, then  
  (\ref{extremum principle}) is  a variational principle for static black holes: it
identifies a  static  black hole configuration  as an  extremum of  total energy, within a  family of   black holes   that are initially at rest.  That is, a   configuration with extremal energy remains at rest, while all other configurations  evolve dynamically.     
  Such  variational principles have been proved for  a
  single uncharged black hole      \cite{Hawking-vp},    a
  single
      black hole  in Einstein-Yang-Mills theory \cite{Sudarsky-Wald, *Chrusciel-Wald}, and     pairs of mirror-symmetric black holes   
  in the   Randall-Sundrum braneworld models  \cite{RS-first-law}.
Our extremum principle (\ref{extremum principle}) adds the case of  
charged black holes  to these earlier variational principles  \cite{Hawking-vp, Sudarsky-Wald, *Chrusciel-Wald,RS-first-law}
and
   identifies the  static  Hartle-Hawking solution as its energy extremum.
In contrast to  \cite{Hawking-vp, Sudarsky-Wald, *Chrusciel-Wald,RS-first-law}, in this paper, the number $N$ of black holes that we consider is   arbitrary, and our energy extremum varies (rather than holds   fixed) the black hole areas.

We     also show that the   energy extremum in (\ref{extremum principle})  is an energy minimum.
This result is consistent with  the earlier result of Gibbons and Hull \cite{Gibbons-Hull}, but our methods  are different.
In \cite{Gibbons-Hull}, it was  found  that the Majumdar-Papapetrou solution     saturates the lower bound  of a 
supersymmetric Bogomol'nyi-Prasad-Sommerfield (BPS)
  inequality,  $E \ge |Q|$, for   \textit{global} quantities: the total energy  $E$ and charge $Q$ (with zero    magnetic charge    in the context of this paper).  In contrast to  \cite{Gibbons-Hull},  in this paper,
we minimize the energy $E(A_i,q_i,r_{ij})$  as a function of several variables,  and we do  not use   supersymmetry; we also  obtain additional extremum conditions ($m_i=|q_i|$)   on   the  \textit{individual}  masses and  charges.

It is worth noting that,
although   black holes with   extremal charge are not often considered  in astrophysical applications,
they
have attracted significant theoretical interest. For example,        they represent stable ground states
for black holes in   supersymmetric theories \cite{supergravity} and they have   been used to investigate the nature of   entropy in black hole thermodynamics \cite{entropy}.
  
This paper is organized as follows.  In Sec.\ \ref{geometry review}, we review  the necessary geometry.  In Sec.\ \ref{extremum}, we prove the extremum principle (\ref{extremum principle})   for   well-separated black holes, through second order, and  we specify    the physical conditions under which the separations $r_{ij}$ are sufficiently large.  In Sec.\ \ref{energy minimum}, we show that the energy extremum is a minimum, and verify that it  agrees with the BPS bound.
  We conclude   in Sec.\ \ref{conclusion}. Throughout this paper, we  work in four spacetime dimensions and use geometric units  with $G=c=4\pi\epsilon_0=1$.

Much of our analysis does not require    that  the black hole charges have the same sign.   We    only need to refer to  same-sign charges in the following contexts:  in  a higher-order analysis (Sec.\ \ref{higher orders}), in   comparisons     to  the BPS   bound (Sec.\ \ref{energy minimum}),  and in applications of   the extremum principle (\ref{extremum principle}) as a variational principle that reproduces  the static Hartle-Hawking solution (as  described above).

\section{GEOMETRY  \label{geometry review}}

In this  section, we review the   geometry   \cite{Brill-Lindquist, Hartle-Hawking} that we    use  in this paper.
A   system of $N$ charged black holes, all initially at rest, is described by its instantaneous spatial geometry.  
The appropriate area $A_i$ of a black hole is that of its  
 apparent horizon, which 
  is   determined   by the  
  spatial geometry alone (unlike   the   event horizon, which is a global  spacetime property).
  The apparent horizon
 generally lies inside  the event horizon, and  coincides with it
  for a static or stationary  spacetime.
  
\begin{figure}
\centering
\includegraphics{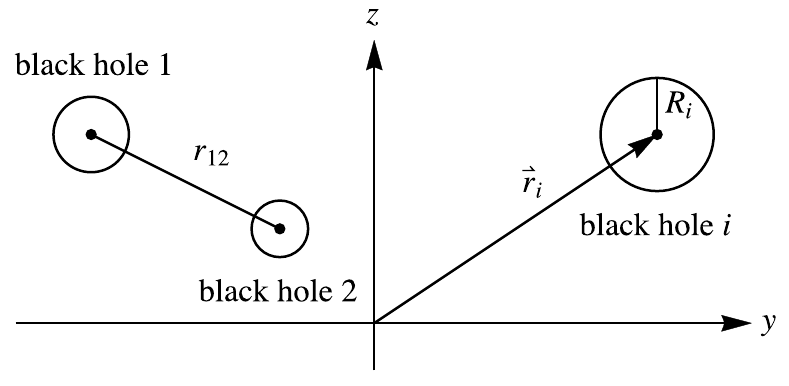}
\caption{\label{figure-geometry}   Illustration of three  nonextremal black holes  in the spherical approximation.  Black holes 1 and 2  are  separated by distance $r_{12}$.  Black hole $i$ has position $\vec{r}_i$ and    apparent horizon   radius $R_i$.   Dots denote the punctures  described in the text.   } 
\end{figure}

We     use  a conformally flat geometry, in the form introduced by Brill and Lindquist \cite{Brill-Lindquist}.
This geometry  contains no conical singularities (struts) to prevent the black holes from moving.
The extremal black holes  considered by Hartle and Hawking    \cite{Hartle-Hawking} remain eternally static.
The nonextremal black holes considered by Brill and Lindquist  \cite{Brill-Lindquist}  are   initially at rest, and evolve  dynamically thereafter  \cite{charged-AH, new-uniqueness}. For both cases, the instantaneous spatial geometry, exterior to    all  $N$ black holes, is  
 \begin{subequations}
 \label{geometry}
\begin{eqnarray}
ds^2 &=& f^2 \left(dx^2 + dy^2 + dz^2\right),
\\*
f &=& \left(1 + \sum_{i=1}^N \frac{\alpha_i}{|\vec r - \vec{r}_i |}\right)
\left( 1 + \sum_{i=1}^N \frac{\beta_i}{|\vec r - \vec{r}_i |}\right).
\end{eqnarray}
\end{subequations}
Figure \ref{figure-geometry}  illustrates the setup.
The Euclidean vector $\vec{r}$  locates any point outside the black holes.   The vector $\vec{r}_i$ locates   black hole   $i$.   
We let $r_{ij}$ denote  a  Euclidean separation distance between black holes $i$ and  $j$,
\begin{equation}
r_{ij} =  |\vec{r}_i - \vec{r}_j |   .
\end{equation}
The parameters $\alpha_i $ and $\beta_i$ are non-negative constants, and are  related to  physical quantities (mass, charge, energy) as follows \cite{Brill-Lindquist, Hartle-Hawking}.  
Black hole   $i$ has  individual mass (rest energy)  $m_i$   and charge $q_i$,
\begin{eqnarray}
\label{mass i}
m_i & =& \alpha_i + \beta_i + \sum_{j \neq i}\frac{(\alpha_i\beta_j + \alpha_j\beta_i)}{r_{ij}} ,
\\*
\label{charge i}
q_i &=& \beta_i - \alpha_i + \sum_{j \neq i} \frac{(\beta_i\alpha_j - \beta_j\alpha_i)}{r_{ij}}   .
\end{eqnarray}
Each pair $(m_i, q_i)$   satisfies the black hole charge-mass inequality, $|q_i| \le m_i$.   
The  system of $N$ black holes has total energy $E$  and  
    interaction energy $E_{\rm int}$,   
\begin{eqnarray}
\label{total energy}
E  &=&   \sum_{i=1}^N (\alpha_i + \beta_i),
\\*
\label{interaction E}
E_{\rm int} &=& E - \sum_{i=1}^N m_i  =
 - \sum_{i=1}^N \sum_{j \neq i} \frac{(\alpha_i\beta_j + \alpha_j\beta_i)}{r_{ij}}  .
\end{eqnarray}

The expressions  for $m_i$ and $q_i$, in (\ref{mass i}) and (\ref{charge i}),  can be solved   for the parameters $\alpha_i $ and $\beta_i$.   For large $r_{ij}$,   the results      to first order  in $1/r_{ij}$ are \cite{Brill-Lindquist} 
 \begin{subequations}
 \label{AB expansion 1}
\begin{eqnarray}
\alpha_i &\simeq & \frac{(m_i-q_i)}{2}\left[1-\frac{1}{2}\sum_{j \neq i} \frac{(m_j+q_j)}{r_{ij}} \right] ,
\\*
\beta_i &\simeq & \frac{(m_i+q_i)}{2}\left[1-\frac{1}{2}\sum_{j \neq i} \frac{(m_j-q_j)}{r_{ij}} \right]  .
\end{eqnarray}
 \end{subequations}
Using (\ref{AB expansion 1}), for large $r_{ij}$, the interaction energy (\ref{interaction E}) can be written      to first order  in $1/r_{ij}$ as
 \cite{Brill-Lindquist}
\begin{equation}
\label{interaction energy}
E_{\rm int}  \simeq  - \sum_{i=1}^N \sum_{j > i} \frac{(m_im_j-q_i q_j)}{r_{ij}}  .
\end{equation}
This  has  the   expected form: it is  the sum of   the pairwise      gravitational  and electrostatic potential energies.

For the static Hartle-Hawking  solution  \cite{Hartle-Hawking},  only one set of the  parameters  $\alpha_i$ and $\beta_i$ are nonzero,
\begin{equation}
\label{HH-A-B}
(\alpha_i =0, \beta_i > 0) \qquad \mbox{or} \qquad (\beta_i =0, \alpha_i > 0) .
\end{equation}
In terms of    mass and charge, the  two cases in (\ref{HH-A-B}) are equivalent to, respectively  \cite{Hartle-Hawking} 
\begin{equation}
\label{HH-m-q}
q_i =m_i  =\beta_i > 0 \qquad \mbox{or} \qquad -q_i=m_i = \alpha_i >0 .
\end{equation}
Thus, the static condition,  (\ref{HH-A-B}) and (\ref{HH-m-q}),  can be summarized as an extremal   condition for same-sign charges,
\begin{equation}
\label{static condition}
|q_i|=m_i, \qquad \mbox{with  all   $q_i$ of like sign.}  
\end{equation}

The black hole charges are   nonextremal, 
if   the parameters  $\alpha_i$ and $\beta_i$  do not satisfy the static condition 
 (\ref{HH-A-B})\textendash(\ref{static condition}).
Near   a nonextremal black hole $i$,
 we may transform from    Cartesian  coordinates $(x,y,z)$ in (\ref{geometry}) to spherical coordinates $(r, \theta,\phi)$ centered at  $\vec{r}_i$.
As found in   \cite{Brill-Lindquist}, for sufficiently  large separations $r_{ij}$,    the   apparent horizon of black hole $i$  can  be treated as
  a sphere,   $r(\theta,\phi) = R_i=$ constant.
 We will  refer to this as the spherical approximation,  which is illustrated  in Fig.\ \ref{figure-geometry}.  The radius $R_i$ is
 given by \cite{Brill-Lindquist}
 \begin{equation}
 \label{r0}
R_i{}^2 =  \alpha_i\beta_i \left(1 +\sum_{j \neq i} \frac{\alpha_j}{r_{ij}}\right)^{-1} \left(1 + \sum_{j \neq i} \frac{\beta_j}{r_{ij}}\right)^{-1}. 
\end{equation}
In the spherical approximation,  the    metric function $f$ in (\ref{geometry}) evaluated on the surface $r=R_i$  is  \cite{Brill-Lindquist}
\begin{equation}
\label{f first order}
f =\left(1 +  \frac{\alpha_i}{R_i} + \sum_{j \neq i} \frac{\alpha_j }{ r_{ij}}   \right)
\left(1 +  \frac{\beta_i}{R_i} + \sum_{j \neq i} \frac{\beta_j }{ r_{ij}}   \right).
\end{equation}
As we   see in Sec.\ \ref{extremum}, the results (\ref{r0}) and (\ref{f first order}) will only hold to up to an appropriate order,
when expanded in powers of the inverse    separations ($1/r_{ij}$).  

We end this subsection by reviewing some technical  features  \cite{Brill-Lindquist, Hartle-Hawking}  of the geometry and topology, for both nonextremal black holes and  extremal black holes.
The  spatial geometry has the  topology  of $\mathbb{R}^3$   with $N$ points  (`punctures') removed. 
Each puncture  is located by the vector  $\vec{r}_i$.  In the      flat background metric, the distance between  two punctures  is the Euclidean distance $r_{ij}$.  

In the full conformal geometry, for  a nonextremal black hole, each puncture  represents  the    spatial infinity of an asymptotically flat region (`sheet'), which is hidden behind the black hole's apparent horizon, as viewed in the common asymptotic region (sheet) exterior to all $N$ black holes, where the total energy $E$ is defined.  The topology  referred to above is equivalent to  the topology  of    $N+1$ connected sheets.  Each black hole's individual mass $m_i$ is the 
Arnowitt-Deser-Misner (ADM) mass of its hidden asymptotic region (its individual sheet). 

  In contrast, 
for an extremal black hole, 
 a careful coordinate analysis \cite{Hartle-Hawking} reveals that the puncture at $\vec{r}_i$ represents the apparent horizon, with finite area.
In this case, the parameter values (\ref{HH-A-B})   cannot be used to evaluate  
  the apparent horizon   radius  (\ref{r0}).
  As shown in  \cite{Hartle-Hawking}, for extremal charge, $|\vec r - \vec{r}_i | \rightarrow 0$  is a simple coordinate singularity,  
  and
  the area of the apparent horizon
 in this limit is  nonzero, $A_i=4\pi m_i{}^2 = 4\pi q_i{}^2$.    This coordinate issue will pose no complications in our proof of the
  extremum principle in Sec.\ \ref{outline of proof},
where     
we will perform all calculations  in the nonextremal regime, and then express the energy $E$ as function of   the finite areas $A_i$.  These areas   reproduce    the  correct values $A_i=4\pi m_i{}^2=4\pi q_i{}^2$   when we take the extremal limit in terms of charges and masses ($|q_i| \rightarrow m_i$), which is a coordinate-independent limit.

\section{ENERGY  EXTREMUM PRINCIPLE \label{extremum}}

\subsection{Physical motivation  \label{motivation}}

In this subsection, we illustrate how  the first law of black hole mechanics motivates
   our extremum principle  (\ref{extremum principle}).  This also  demonstrates  the  basic extremum procedure that we   use in our  proof of 
   the extremum principle 
   (Secs.\ \ref{outline of proof}\textendash \ref{higher orders}).

For clarity, we    begin with the      case  of a single black hole,    
   and then generalize it  to the 
  multi-black hole case.   
A single static charged black hole (the Reissner-Nordstr\"{o}m solution) has 
  energy $E=m$ and charge $|q| \le m$. 
Variations   conserve energy via the  
  first law of black hole mechanics  \cite{Bardeen-four-laws},
  \begin{equation}
  \label{first law 1}
\delta E = \frac{\kappa}{8\pi} \delta A + \Phi \delta q .
\end{equation}
Here $A$,
 $\kappa$, and $\Phi$ are the black hole's surface area, surface gravity, and  electric potential, respectively.
In the case of extremal charge ($|q|=m$),  it is well known that $\kappa=0$, so the first law reduces to
\begin{equation}
  \label{first law 2}
\delta E =   \Phi \delta q.
\end{equation}
The key point of  (\ref{first law 2}) is the following: if $|q|=m$, then   variations  that hold    $q$ constant ($\delta q=0$) extremize   the energy 
($\delta E=0$), and  this means $\partial E/\partial A=0$, since $E$ is  a function of $A$ and $q$.
Conversely, if $q$ is held constant in   (\ref{first law 1}), then an energy extremum requires 
$\partial E/\partial A=0$, and this      reduces to   $|q|=m$, as follows.  The total energy of the
Reissner-Nordstr\"{o}m solution is
   \begin{equation}
   \label{E RN}
  E  =   
  \sqrt{\frac{\pi}{A}}\left(\frac{A}{4\pi} + q^2 \right).
\end{equation}
Evaluating  
  $\partial E/\partial A=0$  for (\ref{E RN}) 
  yields
  $|q|=\sqrt{A/4\pi}$,  for which $E=m=\sqrt{A/4\pi}=|q|$  and hence $|q|=m$.
In summary,   the extremum $\partial E/\partial A=0$ occurs  if and only if  the  extremal charge property holds,  $|q|=m$.

As already noted, the vanishing coefficient of $\delta A$   in the   first law (\ref{first law 2})  motivates why,  for an arbitrarily charged black hole,   we extremize the energy   (\ref{E RN}) by varying the area $A$ and holding  the charge $q$  constant.  This generalizes to  the multi-black hole case, as follows.
The    expected     first law for the static Hartle-Hawking solution  \cite{Hartle-Hawking} is
   \begin{equation}
  \delta E  = \sum_{i=1}^N \Phi_i \delta q_i  ,
\end{equation}
where  $E$ is the total energy. 
For arbitrarily charged black holes, in analogy with the single-black hole case, we extremize $E$ by varying the black hole areas  $A_i$ while holding constant all other quantities: the charges $q_i$ (which appear in the first law)  and      also   the black hole separations $r_{ij}$ (which are internal parameters of the geometry).
This    motivates  our   extremum principle
(\ref{extremum principle}).

Note  that the  single black hole   considered above   remains at rest, so it    illustrates our   energy extremum procedure, but not any   subsequent dynamics.  In this paper, the     initially static    multi-black hole configurations   that are nonextremal  will not remain static; their  dynamics were studied in \cite{charged-AH}.

\subsection{Outline of proof \label{outline of proof}}

In     Secs.\ \ref{first order proof}\textendash \ref{higher orders} below, we will
prove the extremum principle (\ref{extremum principle}) as  an expansion in the inverse    separations, $1/r_{ij}$. 
Zeroth order     ($1/r_{ij} \rightarrow 0$) corresponds to  $r_{ij} \rightarrow \infty$, and treats each black hole 
in isolation from the  other black holes.
Through first order, we retain all terms linear in $1/r_{ij}$, which characterize effectively Newtonian interactions among the black holes. Through second order, we retain all terms quadratic in $1/r_{ij}$, which characterize post-Newtonian relativistic interactions.  

All of the terms in our expansions
 will be
dimensionless ratios  of   black hole properties
($\alpha_i$, $\beta_i$, $m_i$, $q_i$)   divided by the separations $r_{ij}$.  These ratios are small  for sufficiently large separations.  
In particular, we will show that 
the extremum principle   holds for   sufficiently well separated black holes, as specified by    conditions  on $r_{ij}$ that take the form $U_i > 0$, where $U_i$ can be interpreted as  the effective gravitational   potential  experienced by   black hole $i$.

At each order, the main  steps are  to   calculate each black hole's apparent horizon area $A_i$, and then to 
  obtain the  energy $E(A_i,q_i,r_{ij})$ as a function of   the areas, charges, and separations.   We  then extremize the energy 
as  $\partial E/\partial A_i=0$, and from this obtain the condition $|q_i|=m_i$, which proves the extremum principle.

\subsection{Proof through first order \label{first order proof}}

We   begin by calculating   the apparent horizon area $A_i$ of each  black hole.
As in   the original work of Brill and Lindquist \cite{Brill-Lindquist},
through  first order in  $1/r_{ij}$, we use the spherical approximation (see Sec. \ref{geometry review}), for which
  the apparent horizon
is the surface $r=R_i=$ constant in 
    spherical coordinates $(r, \theta,\phi)$   centered at  $\vec{r}_i$.
Then the differential area  is
\begin{equation}
\label{surface metric}
dA_i = 
f^2
   R_i{}^2  \sin^2\theta \, d\theta  \, d\phi    ,
    \end{equation}
    where $R_i$ and  $f$   are    given by
  (\ref{r0}) and (\ref{f first order}), respectively.
  Integrating  then yields the   area formula:
\begin{equation}
\label{area}
A_i = 4\pi {R_i}^2 f^2 .
\end{equation}
We now evaluate this using
  (\ref{r0}) and (\ref{f first order}).  Expanding through first order in $1/r_{ij}$ gives
\begin{equation}
\sqrt{\frac{A_i}{4\pi}}  =   m_i 
+  \sqrt{\alpha_i \beta_i} \left(2+   \sum_{j\neq i} \frac{\alpha_j+\beta_j}{r_{ij}}\right),
\end{equation}
with  
  $m_i$ given by (\ref{mass i}) as a function of the other parameters $(\alpha_i, \beta_i, r_{ij})$.
We can  express the area $A_i$   completely in terms of  the mass $m_i$ and charge  $q_i$,
by substituting $\alpha_i$ and $\beta_i$   in  (\ref{AB expansion 1}).
Then simplifying yields   
\begin{eqnarray}
\label{terms will cancel}
\nonumber
\sqrt{\frac{A_i}{4\pi}} &=& m_i +  \sqrt{m_{i}{}^2 - q_{i}{}^2} 
\\*
& &
\times
\left(1 -   \sum_{j \neq i} \frac{m_j}{2 r_{ij}} \right)
\left(1 + \sum_{j \neq i} \frac{m_j}{2r_{ij}} \right) .
\end{eqnarray}
When we expand this expression, the  terms that are first order in   $1/r_{ij}$ exactly cancel. At first order, we neglect  the terms quadratic in $1/r_{ij}$, and obtain
\begin{equation}
\label{area first order}
\sqrt{\frac{A_i}{4\pi}} = m_i + \sqrt{m_i{}^2 - q_i{}^2}   .
\end{equation}
 Solving this   for the mass $m_i$ yields 
\begin{equation}
\label{mass i leading order}
m_i = \sqrt{\frac{\pi}{A_i}}\left(\frac{A_i}{4\pi} + q_i{}^2 \right)   .
\end{equation}
Due to the cancellation that occurred to obtain  
(\ref{area first order}),
the mass    (\ref{mass i leading order})  contains neither first-order terms (proportional to  $1/r_{ij}$)  nor any parameters related to the other black holes ($j \neq i$).  This
first-order   result  is  therefore the same as the zeroth-order result, and it has the same form as the energy (\ref{E RN}) for a single black hole.

We now evaluate the total energy $E$, which  by    (\ref{interaction E}) and (\ref{interaction energy}) is, to first order in $1/r_{ij}$,
\begin{equation}
\label{E 1}
E = \sum_{i = 1}^N m_i - \sum_{i = 1}^N \sum_{j > i}\frac{( m_i m_j - q_i q_j)}{r_{ij}}   .
\end{equation}
The zeroth-order term  in (\ref{E 1}) is the sum of  the    rest energies, and the first-order terms are the  pairwise    gravitational and  electrostatic  potential energies;    
this   will be useful in interpreting the quantity $U_i$ below. 

We  now take each mass $m_i$   in (\ref{E 1}) to be given by  (\ref{mass i leading order}), which expresses the  energy  (\ref{E 1}) in the functional form  $E(A_i,q_i,r_{ij})$.  
Then holding   constant the charges $q_i$ and separations $r_{ij}$,
we find
\begin{equation}
\label{dE 1}
  \frac{\partial E}{\partial A_i} = U_i  \,\frac{\partial m_i}{\partial A_i} ,
\end{equation}
where
\begin{equation}
\label{U 1}
U_i =    1 -  \sum_{j \neq i} \frac{m_j}{r_{ij}}   .
\end{equation} 
 We  now extremize the  total energy $E$ by requiring $\partial E/\partial A_i=0$, which by  (\ref{dE 1})    is equivalent to 
 \begin{equation}
\label{dE  extremum 1}
 U_i \,\frac{\partial m_i}{\partial A_i} = 0 .
\end{equation}
  For sufficiently large separations,   $U_i \neq 0$, as  seen by  (\ref{U 1}).
   In this case,  the   extremum (\ref{dE  extremum 1})      requires
\begin{equation}
\label{dm 1}
\frac{\partial{m_i}}{\partial{A_i}} = 0  .
\end{equation}
 Evaluating this for (\ref{mass i leading order}) reduces to   
\begin{equation}
\label{q zeroth}
 |{q_i}| =  \sqrt{\frac{A_i}{4\pi}}   .
\end{equation}
 For this charge value,   the mass (\ref{mass i leading order}) is
\begin{equation}
\label{mass zeroth}
  m_i =     \sqrt{\frac{A_i}{4\pi}}  = |q_i| .
\end{equation}
Thus, since the     extremum (\ref{dE  extremum 1})    has produced the  $N$ desired conditions, $|q_i|=m_i$,   we have  proved the extremum principle  (\ref{extremum principle})  through first  order.

As noted above, our proof  requires   $U_i \neq 0$.   As seen by  (\ref{U 1}), this  condition  
is always satisfied at zeroth order   ($1/r_{ij}\rightarrow 0$ and $U_i=1$), and  it is   satisfied  at first order
for sufficiently large separations $r_{ij}$ compared to the masses $m_i$.
   In particular, the continuity
of our perturbative approach with the zeroth-order limit 
($U_i=1$) requires
  $0 < U_i \le 1$.

We end this subsection by interpreting the quantity $U_i$.
Note that $U_i$ can be obtained  from     the energy    (\ref{E 1})  in the form $E(m_i,q_i,r_{ij})$, along with (\ref{dE 1}) and the chain rule,    
\begin{equation}
\label{U 1 fund}
U_i =  \frac{\partial E}{\partial m_i}   .
\end{equation}
This  expression motivates the interpretation of $U_i$     
 as   the effective   gravitational 
 potential experienced by  black hole $i$, including its   rest energy.
 This interpretation is justified  by   the derivation of (\ref{dE 1})\textendash(\ref{U 1}) from (\ref{E 1}), which makes clear that: (i)
the zeroth order term in  (\ref{U 1})  is  the   rest energy, normalized per unit mass; (ii) the first-order sum in  (\ref{U 1}) is the Newtonian  gravitational potential;
 (iii) there is no electric potential contribution      to $U_i$  since (\ref{dE 1})\textendash(\ref{U 1}) are obtained by holding   the charges  constant in (\ref{E 1}).

\subsection{Proof through second order \label{second order proof}}

 \begin{figure}
\centering
\includegraphics{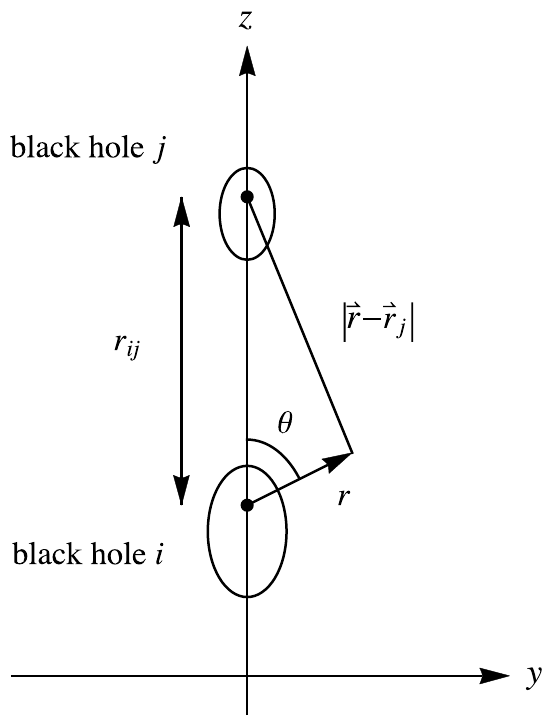}
\caption{\label{figure-second-order}   Illustration of two    nonextremal   black holes  in the  second-order analysis, showing useful distances and    coordinates  $(r,\theta)$. The dots have  Euclidean separation $r_{ij}=|{\vec r}_i -{\vec r}_j|$. All black holes (including  others, not shown) are collinear on the   $z$ axis, and have azimuthal symmetry about the $z$ axis.   
} 
\end{figure}

In this section, we   prove the extremum principle  (\ref{extremum principle}) through second order.  
Our first task  will be to evaluate each  apparent horizon   area   $A_i$.   This    is a longer calculation  than in our first-order proof  (Sec.\ \ref{first order proof}), since at second order,  we       must  depart from the spherical approximation.  
Figure \ref{figure-second-order} illustrates  the   setup.

In general, each  black hole's apparent horizon   is  nonspherical, due to its interactions with  the other black holes.  
Thus, in
   spherical coordinates $(r,\theta,\phi)$ centered at    ${\vec r}_i$,
 the apparent horizon  of
black hole $i$ is generally  a  surface $r(\theta,\phi)$.
     To simplify the analysis, we   will  take   
all black holes to be    collinear (aligned on the   $z$ axis). The resulting axisymmetry simplifies each nonspherical   horizon $r(\theta,\phi)$  to the    azimuthally symmetric form $r(\theta)$.
Thus, as
  illustrated in Fig.\ \ref{figure-second-order},
we take all  $N$ black hole positions ${\vec r}_i$    to lie   on the $z$ axis. 
Near black hole $i$, we introduce  spherical coordinates $(r,\theta,\phi)$ centered at  ${\vec r}_i$,  and
consider another black hole labeled by $j \neq i$.  This gives
\begin{subequations}
\label{collinear}
\begin{eqnarray}
|\vec r - \vec{r}_i | &=& r ,
\\*
\label{distance r-rj}
|\vec r - \vec{r}_j |   &=& \sqrt{r^2 - 2r_{ij} \cos\theta +r_{ij}{}^2}.
\end{eqnarray}
\end{subequations}
To calculate the area $A_i$,  we    only need  to consider     points $(r,\theta)$    on the apparent horizon,   so    $r<r_{ij}$.
In this case,    the    distance   (\ref{distance r-rj})  has a standard expansion in Legendre polynomials $P_k$,
\begin{equation}
\label{Legendre expansion}
\frac{1}{|\vec r - \vec{r}_j |}  = \frac{1}{r_{ij}} \sum_{k \ge 0} \left(\frac{r}{r_{ij}}\right)^k P_k(\cos\theta).
\end{equation}
The form (\ref{Legendre expansion}) is very suitable for our
purposes, since it is an expansion in the inverse separations $(1/r_{ij})$.
Our first-order proof (Sec.\ \ref{first order proof}), like  the original  work of Brill and Lindquist   \cite{Brill-Lindquist},  used the summation  in   (\ref{Legendre expansion}) through $k=0$ and $P_0=1$.  Here, our second-order proof proceeds through  $k=1$ and $P_1=\cos\theta$.

We now use   (\ref{collinear}) and (\ref{Legendre expansion})  to evaluate    the metric function $f$ in (\ref{geometry}) near black hole $i$.  We retain all terms through second order in  $1/r_{ij}$.  This gives
\begin{eqnarray} 
\nonumber
f(r,\theta) &=& 1 + \frac{\alpha_i+\beta_i}{r}  + \frac{\alpha_i \beta_i}{r^2}
       +    \sum_{j \neq i} \frac{\alpha_j }{r_{ij}}  \sum_{k \neq i} \frac{\beta_k }{r_{ik}} 
\\*
\nonumber
   & & 
   + 
   \sum_{j\neq i}
   \left[
    \left( \frac{1}{r_{ij}}+ \frac{r\cos\theta}{r_{ij}{}^2}  \right) 
  \right.
  \\*
  & &
    \label{f(r,t)}
\qquad    \times
    \left.
  \left( \alpha_j+\beta_j  +   \frac{\alpha_i \beta_j + \beta_i \alpha_j}{r}  \right)
     \right] .
\end{eqnarray}
This contains  all of the second-order terms from   expanding  the  spherical approximation (\ref{f first order}), and generalizes it by including nonspherical  terms, proportional to $\cos\theta$.

 Each black hole's  apparent horizon $r(\theta)$ 
is an extremal surface, so it  
is determined by   extremizing   the area   \cite{Brill-Lindquist},  
\begin{equation}
\label{area functional}
 A_i  = 2\pi \int_0^\pi d\theta\
\sin\theta  \, r
\sqrt{r^2+ \dot{r}^2}\, f^2   ,
\end{equation}
where $\dot r = dr/d\theta$.  Extremizing  (\ref{area functional}) gives 
\begin{equation}
\label{Lagrange r}
\frac{r+\ddot{r}}{1+\dot{r}^2/r^2} =
-  \dot{r}
\left(
\cot\theta +  \frac{2\partial_\theta f}{f}
\right)
 + r \left( 3 +   \frac{2r\partial_r f}{f}\right)  .
\end{equation}
As in \cite{Brill-Lindquist}, we will solve
(\ref{Lagrange r}) by expanding in   Legendre polynomials $P_n$  as  basis functions.  This    function-basis  expansion
for $r(\theta)$ is   distinct from  our previous  multipole  expansion 
(\ref{Legendre expansion}).
Thus, we take
\begin{equation}
\label{basis r}
r(\theta)=  \sum_{n \ge 0} C_{ni} P_n(\cos\theta),
\end{equation}
where $C_{ni}$  are constant coefficients and   $i$ labels the black hole.  
Our first-order proof (Sec.\ \ref{first order proof}), like  the original  work of Brill and Lindquist   \cite{Brill-Lindquist}, used the expansion  (\ref{basis r}) through $n=0$, which is the spherical approximation (\ref{r0}).  
At second order, 
 we use  (\ref{basis r})  through $n=1$,
\begin{equation}
\label{r(t)}
r(\theta)=   C_{0i}+ C_{1i} \cos\theta ,
\end{equation}
with $P_0=1$ and $P_1=\cos\theta$.
To determine the coefficients  $C_{ni}$,
 we evaluate the differential equation  (\ref{Lagrange r}) for $r(\theta)$
using    (\ref{r(t)})  and the metric function  $f$ in (\ref{f(r,t)}). After   expanding all quantities to second order in  
   $1/r_{ij}$, (\ref{Lagrange r}) then takes the form
\begin{equation}
\label{Lagrange reduced}
F_{0} P_0 + F_{1} P_1(\cos\theta) = 0, 
\end{equation}
where $F_0$ and $F_1$ are  functions (which  for brevity we do  not list here) that are independent of $\theta$, but involve the other 
parameters ($\alpha_i$, $\beta_i$, $r_{ij}$).
Since (\ref{Lagrange reduced}) is an expansion in the basis functions $P_n$, both $F_0$ and $F_1$ must vanish.
   Solving the equation $F_0=0$   reduces to  
 \begin{equation}
 C_{0i} = R_i ,
  \label{coeff C0}
\end{equation}
where $R_i$  is the    radius in the spherical approximation (\ref{r0}),        which is here to be expanded through second order in $1/r_{ij}$.  Solving the equation $F_1=0$   yields
\begin{eqnarray}
\nonumber
C_{1i}  &=&  
- \left( \frac{\alpha_i \beta_i}{4\sqrt{\alpha_i \beta_i}  + \alpha_i+\beta_i} \right)
\\*
& & 
\times
 \sum_{j\neq i}\frac{2\sqrt{\alpha_i \beta_i} (\alpha_j+\beta_j)+\alpha_i \beta_j + \beta_i \alpha_j }{r_{ij}{}^2}
 .
 \label{coeff C1}
\end{eqnarray}
Thus, the coefficients $C_{1i}$ are  second-order quantities.
In \cite{Brill-Lindquist}, the coefficients $C_{ni}$  were determined numerically for uncharged black holes, and $|C_{ni}|$  was found to decrease as $n$ increases.  Our results, (\ref{coeff C0}) and (\ref{coeff C1}),  show analytically
 that here
 $|C_{ni}|$   decreases as $n$ increases
  due to its      leading power in the  inverse separations, $(1/r_{ij})^{2n}$.

For the  coefficients $ C_{ni}$  in (\ref{coeff C0}) and (\ref{coeff C1}), the        horizon surface 
  (\ref{r(t)})      approximates   a  prolate     ellipsoid  with one focus at $r=0$.  This can be seen by linearizing the   surface function $r_{\rm e}(\theta)$ of   a prolate  ellipsoid    in the eccentricity ($\epsilon \ll 1$),  
\begin{equation}
r_{\rm e}(\theta) = \frac{a(1- \epsilon^2)}{1+\epsilon \cos\theta} 
\simeq a-a\epsilon \cos\theta .
\end{equation}
The linearized form matches    the  horizon (\ref{r(t)}), so the horizon approximates  an ellipsoid with semimajor axis   $a=C_{0i}=R_i$ and   eccentricity    $\epsilon = C_{1i}/R_i$, as illustrated in  Fig.\ \ref{figure-second-order}.

We now evaluate the horizon area $A_i$ in
(\ref{area functional}) for the surface $r(\theta)$ in (\ref{r(t)}),
 using the metric function  $f$ in (\ref{f(r,t)}). Expanding through second order in $1/r_{ij}$, we find
\begin{eqnarray}
\nonumber
\sqrt{\frac{A_i}{4\pi}} &=&  m_i 
+  \sqrt{\alpha_i \beta_i} \left(2+   \sum_{j\neq i} \frac{\alpha_j+\beta_j}{r_{ij}}\right)
\\*
& & 
\label{area AB 2}
-\,\frac{\sqrt{\alpha_i \beta_i}}{4}\left(\sum_{j\neq i}\frac{\alpha_j-\beta_j}{r_{ij}}\right)^2 ,
\end{eqnarray}
with   $m_i$  given by (\ref{mass i}) as a function of the other parameters    $(\alpha_i, \beta_i, r_{ij})$.
The result   (\ref{area AB 2})  is   the first-order area (\ref{area first order}), plus second-order corrections.

Our next step is to obtain the  energy $E(A_i,q_i,r_{ij})$ as a function of   the areas, charges, and separations.    
We   evaluate the energy $E$ by a different method   than in our first-order proof (Sec.\ \ref{first order proof}), where we used the  Newtonian form  (\ref{E 1}) of the energy,
and expressed the   masses $m_i$  
 in terms of the areas $A_i$. 
At second order, our  setup is    post-Newtonian,
and we find it convenient to  
 express  the parameters $\alpha_i$ and $\beta_i$ in terms of the areas $A_i$ and charges $q_i$, and then evaluate
  the energy    (\ref{total energy}) in the form $E  =   \sum_i (\alpha_i + \beta_i)$. 

Thus, we must solve the area equation (\ref{area AB 2}) and the charge definition (\ref{charge i}) for the parameters $(\alpha_i, \beta_i)$
  as functions of   the areas and charges
$(A_i,  q_i)$. 
For convenience in summarizing our results below,  we define
the following functions 
$(\mu_i, {\cal M}_i, {\cal Q}_i)$  of the areas  and charges.
We define the quantity $\mu_i$   as   
 \begin{equation}
\label{mu-def}
\mu_i =  
 \sqrt{\frac{\pi}{A_i}}\left(\frac{A_i}{4\pi} + q_i{}^2 \right),
\end{equation}
which is our   mass result (\ref{mass i leading order}) through first order.
We define the two dimensionless  quantities $({\cal M}_i, {\cal Q}_i)$   as
\begin{equation}
\label{MQ-def}
{\cal M}_i =  \sum_{j\neq i} \frac{\mu_j}{r_{ij}} ,
\qquad
{\cal Q}_i =  \sum_{j\neq i} \frac{q_j}{r_{ij}}  .
\end{equation}
We now solve  for the parameters $(\alpha_i, \beta_i)$  using a   perturbation approach: we write  
$\alpha_i$ and $\beta_i$ as their  leading-order values (\ref{AB expansion 1}), 
 plus second-order corrections $(\widetilde \alpha_i, \widetilde \beta_i)$.   
We rephrase   the leading-order values (\ref{AB expansion 1})  using  the definitions
(\ref{mu-def}) and (\ref{MQ-def}).
This gives
 \begin{subequations}
 \label{AB ansatz}
\begin{eqnarray}
\label{alpha 2 ansatz}
\alpha_i &=& \frac{(\mu_i-q_i)}{2}\left[1-\frac{({\cal M}_i + {\cal Q}_i )}{2}  \right]+ \widetilde \alpha_i ,
\\*
\beta_i &=& \frac{(\mu_i+q_i)}{2}\left[1-\frac{({\cal M}_i - {\cal Q}_i )}{2}  \right] + \widetilde \beta_i .
\end{eqnarray}
\end{subequations}
It remains to solve for    $\widetilde \alpha_i$ and $\widetilde \beta_i$ as
functions of the areas and charges $(A_i,  q_i)$.
To this end, we    insert  
 (\ref{AB ansatz})  into   the    area  result   (\ref{area AB 2}) and into the charge definition   (\ref{charge i}).  After expanding through second order in $1/r_{ij}$,  we  then find that  (\ref{area AB 2})  and (\ref{charge i})  reduce to, respectively,  
\begin{equation}
 \label{AB to solve}
K_+ \widetilde \alpha_i + K_-  \widetilde \beta_i = \varepsilon_i ,
\qquad
\widetilde \beta_i -   \widetilde \alpha_i    = \delta_i.
\end{equation}
We will summarize  the coefficients $K_\pm$ and  source terms  $(\varepsilon_i, \delta_i)$ below.
 Solving the   linear equations  (\ref{AB to solve}) yields the  second-order corrections,  
   \begin{equation}
 \label{AB corrections}
  \widetilde \alpha_i   = \frac{ \varepsilon_i  - K_- \delta_i }{K_+ + K_-} 
, 
\qquad
\widetilde \beta_i  = \frac{ \varepsilon_i  + K_+ \delta_i }{K_+ + K_-}  .
\end{equation} 
The coefficients $K_\pm$ and source terms  $(\varepsilon_i, \delta_i)$ are
 \begin{eqnarray}
 K_\pm &=&  
1 + \frac{\mu_i \pm   q_i}{\sqrt{\mu_i{}^2-q_i{}^2}},
 \\*
\varepsilon_i    
  &=& \sqrt{\frac{A_i}{4\pi}} \, {\cal S}_i - q_i    {\cal T}_i,
\\*
  \delta_i   
&  =& q_i {\cal S}_i   - \mu_i  {\cal T}_i.
\end{eqnarray}
For convenience,  we  have let ${\cal S}_i$ and  ${\cal T}_i$  denote the  following    
dimensionless second-order   quantities,
\begin{subequations}
 \label{ST def}
\begin{eqnarray}
  {\cal S}_i &=& 
\frac{ {\cal M}_i{}^2 +   {\cal Q}_i{}^2 }{4}
 +  \sum_{j\neq i} 
 \sum_{k \neq j }   \frac{(\mu_j\mu_k - q_jq_k)}{4 r_{ij} r_{jk}} ,
 \\*
   {\cal T}_i &=& 
 \frac{ {\cal M}_i {\cal Q}_i  }{2}
    -
 \sum_{j\neq i} 
  \sum_{k \neq j }   \frac{  (\mu_j q_k - q_j\mu_k )}{4 r_{ij} r_{jk}} .
\end{eqnarray}
  \end{subequations}
Two simple results   follow from  using the second-order solutions
 for $\alpha_i$ and $\beta_i$.
First, when we evaluate the      mass $m_i$  in   (\ref{mass i}) using   (\ref{AB ansatz})\textendash(\ref{ST def}),   a significant number of terms cancel, and we find
  \begin{equation}
  \label{m 2}
  m_i   =  \mu_i ,
\end{equation}
with $\mu_i$ defined in (\ref{mu-def}).  Interestingly,  
this is the same as the mass function     (\ref{mass i leading order}) through first order.
Henceforth, we   write $m_i$ in place of $\mu_i$.
A second result is that, after using
(\ref{AB corrections})\textendash(\ref{ST def}),
 the   sum $ \widetilde\alpha_i + \widetilde\beta_i$ 
  reduces to
\begin{equation}
  \label{AB sum corrections}
  \widetilde\alpha_i + \widetilde\beta_i    =
 m_i {\cal S}_i -  q_i  {\cal T}_i .
\end{equation}
 We   now  evaluate  the total energy 
 $E  =   \sum_i (\alpha_i + \beta_i)$,   as given by (\ref{total energy}).
Using  (\ref{AB ansatz}) and   (\ref{AB  sum corrections}) gives
\begin{equation}
\label{E 2}
E = \sum_{i=1}^N
\left[
m_i - \frac{1}{2}( m_i {\cal M}_i - q_i {\cal Q}_i  )   +   m_i {\cal S}_i -  q_i  {\cal T}_i   
\right] .
\end{equation}
The zeroth-order and first-order terms  in  (\ref{E 2}) reproduce  our previous result  (\ref{E 1}), which is the sum of   the rest energies and the pairwise   Newtonian   and electrostatic potential energies.
The second-order terms in  (\ref{E 2}), proportional to ${\cal S}_i$ and ${\cal T}_i$, are relativistic
  post-Newtonian   corrections.   

The energy    (\ref{E 2}) can be regarded as a function  $E(A_i,q_i,r_{ij})$ of the areas, charges, and separations.
This   follows from using   the definitions of the quantities   shown
 $(m_i, {\cal M}_i, {\cal Q}_i,  {\cal S}_i,  {\cal T}_i)$.
  Holding   constant the charges $q_i$ and separations $r_{ij}$, we then find 
\begin{equation}
\label{dE 2}
  \frac{\partial E}{\partial A_i} = U_i \,\frac{\partial m_i}{\partial A_i}, 
\end{equation}
where    
\begin{eqnarray}
\nonumber
U_i &=&
1- {\cal M}_i
+
 \frac{1}{2} \left({\cal M}_i{}^2 +  {\cal Q}_i{}^2\right)
 \\*
 \nonumber
 & & 
+   \frac{3}{4}  \sum_{j\neq i}  \sum_{k \neq j} \frac{(m_j m_k - q_j q_k)}{r_{ij} r_{jk}}
 \\*
 & & 
+   \frac{1}{4}  \sum_{j} \sum_{ k \neq i, j } \frac{(m_j m_k - q_j q_k)}{ r_{ik} r_{jk}}.
\label{U 2}
\end{eqnarray}
In (\ref{U 2}),
the leading  term $(1- {\cal M}_i)$ is our   first-order result   (\ref{U 1}) for $U_i$. The additional  terms  in   (\ref{U 2}), which are second order, are relativistic post-Newtonian     contributions.

The remaining steps in our   proof  are  now          essentially the same    as    (\ref{dE  extremum 1})\textendash(\ref{mass zeroth}) in 
our first-order proof (Sec.\ \ref{first order proof}). 
 We    extremize the  total energy $E$ in (\ref{E 2}) by requiring $\partial E/\partial A_i=0$, which by  (\ref{dE 2})  is equivalent to 
 \begin{equation}
\label{dE  extremum 2}
 U_i \,\frac{\partial m_i}{\partial A_i} = 0 .
\end{equation}
  For sufficiently large separations,   $U_i \neq 0$.  In this case,  the   extremum (\ref{dE  extremum 2})      requires $\partial m_i /\partial A_i =0$.
 Evaluating this  for  (\ref{mu-def}) reduces to     $|{q_i}| =  \sqrt{A_i/4\pi}$.  
 For this charge value,  the  mass     (\ref{mu-def})  is $m_i =  \sqrt{A_i/4\pi} =  |q_i|$.

Thus, since the     extremum (\ref{dE  extremum 2})    has produced the  $N$ 
 desired conditions, $|q_i|=m_i$, 
  we have  proved the extremum principle  (\ref{extremum principle})  through second  order.
As noted above, our proof  requires  $U_i \neq 0$.
This  condition  
is always satisfied at zeroth order   (for which $U_i=1$), and  it is   satisfied  through second order
for sufficiently large separations $r_{ij}$ compared to the masses and charges, as seen by   (\ref{U 2}), and by (\ref{U explicit}) below.
As in the first-order case,  continuity
of our perturbative approach with the zeroth-order limit 
($U_i= 1$) requires
  $0 < U_i \le 1$.

We end this subsection by revisiting the  interpretation of  $U_i$.
As in the first-order case (Sec.\ \ref{first order proof}),  
note that $U_i$ can be obtained  from     the energy  (\ref{E 2})   in the form $E(m_i,q_i,r_{ij})$, along with (\ref{dE 2}) and the chain rule,    
\begin{equation}
\label{U 2 fund}
U_i =  \frac{\partial E}{\partial m_i}   .
\end{equation}
As in the first-order case, this motivates the interpretation of    $U_i$ as the effective   gravitational 
 potential experienced by  black hole $i$, including its rest energy. 
In this interpretation, it is not surprising that  the   gravitational potential   (\ref{U 2})  contains quadratic charge-dependent terms; this is because    nonzero electric stress-energy (a  quadratic function of  the electric field) contributes to curving  the geometry, and so   contributes  to  the  gravitational field.
As an explicit example, for    two black holes ($N=2$),
\begin{equation}
\label{U explicit}
U_1 = 1- \frac{m_2}{r_{12}} + \frac{m_2{}^2 + q_2{}^2 + 2 (m_1m_2 - q_1q_2)}{2r_{12}{}^2} ,
\end{equation}
and  $U_2$ is similarly given by interchanging  all subscripts  ($1 \leftrightarrow 2$).
If we interpret $U_1$ as an 
 effective 
 gravitational  potential, then
 the   terms  proportional to $m_1m_2-q_1q_2$ (which   refer  to black hole 1 itself)
  represent the    nonlinear   gravitational  coupling     in general relativity  between   black hole  1 and the other sources of 
  energy $(m_2, q_2)$.

\subsection{Higher orders \label{higher orders}}

 At higher orders, 
a
  proof  of the extremum principle    (\ref{extremum principle}) 
can be expected to proceed  similarly to Sec.\  \ref{second order proof}, so   this subsection  
 provides additional comments, rather than a full  analysis. 
Through second order, as already noted in  Sec.\ \ref{second order proof},  the mass $m_i$ in (\ref{m 2}) has the same form
as the result    (\ref{mass i leading order}) through first order: it is  the same intrinsic function of the black hole's area  $A_i$ and charge $q_i$, independent of       the other black holes ($j \neq i$).
This result  is   perhaps unexpected, based on  (\ref{mass i}), where   $m_i$ is a summation that   involves      the other black holes $(j \neq i)$.    
However, it could be anticipated physically, since the mass $m_i$   is specific to black hole $i$.  Each mass $m_i$ contributes the the total energy, 
\begin{equation}
\label{E 3}
 E = \sum_{i=1}^N m_i + E_{\rm int} .
\end{equation}
Our goal is to evaluate and extremize the energy function $E(A_i,q_i,r_{ij})$.
From the observations above,  the sum of the masses $m_i$ in (\ref{E 3}) can  be anticipated to take a simple form at higher orders.
To evaluate   $E$, it remains to consider the interaction energy, which we can view   as the essential new task at each higher order,  as a comparison  of (\ref{E 1})  and  (\ref{E 2}) illustrates; the steps  will be similar to those   following the   evaluation of  the area $A_i$  in Sec.\ \ref{second order proof}.

Through second order, our results of Sec.\  \ref{second order proof}  show that  that   $E_{\rm int}=0$ at the energy extremum, if all charges have the same sign. This is seen by letting let $q_i=\pm m_i$ for  extremal charges, each with the same sign  ($\pm$). This   gives 
  ${\cal Q}_i =\pm {\cal M}_i$ 
  and 
   ${\cal T}_i = \pm {\cal S}_i$
     from (\ref{MQ-def}) and  (\ref{ST def}),
        respectively. Then several terms cancel in the energy (\ref{E 2}),  which reduces to
$E = \sum_{i} m_i$,
and  shows that $E_{\rm int}=0$.

It is promising  to note that, at all orders,   the following converse of the above result holds:
if all charges are extremal and have  the same sign, then the interaction energy vanishes, $E_{\rm int}=0$.  This is    most  easily seen by substituting the     extremal charge  condition in the form (\ref{HH-A-B}) into the interaction energy (\ref{interaction E}).   
Since this is an exact result, it must hold  at all orders, when all quantities are expanded in the inverse separations.
The vanishing of $E_{\rm int}$  for   extremal same-sign charges   is an exact supporting result, and   is a precise statement about energy, like the extremum principle.

\section{Energy minimum \label{energy minimum} }

For sufficiently well separated black holes, as specified by the condition
$U_i >0$ (see Secs.\ \ref{first order proof} and \ref{second order proof}),
it is straightforward to show that our energy extremum, 
(\ref{dE  extremum 1})  and (\ref{dE  extremum 2}),
is a minimum.   To show this, we verify that  the  second derivatives of $E$ are positive.
Differentiating
 (\ref{dE 1}) and  (\ref{dE 2}), holding constant the charges $q_i$ and separations $r_{ij}$, gives
\begin{equation}
\label{d2E 1}
  \frac{\partial^2 E}{\partial A_i{}^2} = U_i  \frac{\partial^2 m_i}{\partial A_i{}^2}   
  +   \frac{\partial U_i}{\partial A_i}    \frac{\partial m_i}{\partial A_i}
  .
\end{equation}
We then evaluate  $\partial^2 m_i/\partial A_i{}^2$ from (\ref{mass i leading order}) or (\ref{mu-def}),
and we
evaluate at the extremum ($\partial m_i/\partial A_i=0$).  This gives
\begin{equation}
\label{d2E  extremum 1}
  \frac{\partial^2 E}{\partial A_i{}^2} =  \frac{U_i}{8\sqrt{\pi}A_i{}^{3/2}} ,
\end{equation}
 where $U_i$  is  to be evaluated with $m_i=|q_i|$. The condition   $U_i >0$ 
 ensures that   (\ref{d2E  extremum 1}) is positive,   hence   the energy extremum   is  a   minimum.

If  each charge $q_i$ has the same sign,    it is  also straightforward to verify that  at the extremum,  $E=|Q|$, where  $Q$ is  the total charge.  This is seen as follows.
Let $q_i=\pm m_i$ for  extremal charges, each with the same sign  ($\pm$). 
 Then  
\begin{equation}
 Q = \sum_{i=1}^N q_i = \pm \sum_{i=1}^N  m_i .
 \end{equation}
 As already noted in Sec.\ \ref{higher orders},
 for    extremal   same-sign charges ($q_i=\pm m_i$), we also have    
\begin{equation}
 E = \sum_{i=1}^N m_i = \pm Q  = |Q| .
 \end{equation}
The energy minimum and its extremum value   are both  consistent with the BPS bound in the  energy inequality \cite{Gibbons-Hull}  that is satisfied by the Hartle-Hawking static solution \cite{Hartle-Hawking}.  
 In contrast to \cite{Gibbons-Hull}, we   have    minimized the energy 
 $E(A_i,q_i,r_{ij})$ as a function   of   physical variables,  
 without    using  supersymmetry.

\section{CONCLUSION \label{conclusion}}

Our main result in this paper is  an extremum principle, which derives      the extremal charge condition ($|q_i|=m_i$)  for  a set $N$  black holes (all initially at rest and    arbitrarily charged) by 
extremizing the total energy $E(A_i,q_i,r_{ij})$ with respect to   the black hole horizon areas, at fixed charges and  Euclidean separations.  
  This   principle is motivated by the first law of black hole mechanics, and is valid 
 if the black holes are  sufficiently well separated, as specified by the   condition    
$U_i >  0$, where     $U_i$ can be interpreted as the effective gravitational potential experienced by black hole $i$.  

Our energy extremum is taken at fixed   Euclidean  separations $r_{ij}$, rather than fixed  
   proper distances  between horizons.  This is natural, in the sense that the  proper distances  are well known to become infinite for extremal charges, while the separations $r_{ij}$ remain finite.

If all of the black hole charges have the same sign, then
our extremum principle is a type of variational principle, and augments a list of 
  existing   principles   \cite{Hawking-vp, Sudarsky-Wald, *Chrusciel-Wald,RS-first-law} 
that interpret   static black holes 
(here,
 the static Hartle-Hawking   
     solution)      as  extrema of total energy.  
Our results are  also    consistent with the supersymmetric BPS   energy minimum   ($E=|Q|$) of \cite{Gibbons-Hull}.   Our derivation also shows how the     
 corresponding      substructure  arises   ($m_i=|q_i|$)   for each  individual  black hole.  
 To our knowledge, our results  provide  the first  energy interpretation  of   the static Hartle-Hawking   
     solution \cite{Hartle-Hawking}. This interpretation requires neither the use of balanced forces, nor supersymmetry.

It would be interesting to consider   the regime of very small black hole separations, which is beyond the scope of this paper. This   would  probably require numerical methods, and would   involve  an outermost  common apparent horizon surrounding two   nonextremal black holes that are sufficiently close to  each other,
similar to  the case of uncharged black holes released  at rest \cite{Brill-Lindquist, Bishop}.   
The   formation of a common apparent horizon 
 has   been studied numerically  in the 
head-on collision  of    symmetric like-charged black holes  \cite{charged-AH} for charge-to-mass ratios   in the range $ 0\le q/m \le 0.98$.   

   In this paper,  to locate   a black hole's apparent horizon analytically,  we have done so perturbatively, which  is a well-known feature of the Brill-Lindquist geometry  \cite{Brill-Lindquist} that we have employed.   
 One might wonder if a different geometry   could be used instead; this  appears to be unlikely. For example,  in   the conformally flat   geometry  found from Misner's well-known method of images  \cite{Misner,Lindquist},   
  each black hole's apparent horizon is designed to be an exact coordinate sphere in the flat background space, which is analytically convenient.  This is achieved
by constructing the solution    as an infinite series (similar to  the method of images in   electrostatics).
However, this approach
does not     permit black holes with extremal charge, since the infinite series solution must satisfy  a    convergence condition.  This  condition is rather formal in general; for two symmetric black holes   with opposite charges, it reduces to the statement that the black holes  are nonextremal   \cite{Lindquist}.  A similar restriction   can be expected for  same-sign charges.  
This suggests that the Brill-Lindquist geometry \cite{Brill-Lindquist}  is the unique  family    of nonextremal solutions that  smoothly connects to the extremal Hartle-Hawking solution.

 \begin{acknowledgments}
This work was supported by the  College of Science and Mathematics at   California Polytechnic State University,  
San Luis Obispo.
We   thank Don Spector for   useful    discussions at the Kavli Institute for Theoretical Physics, University of California, Santa Barbara.
\end{acknowledgments}

\bibliography{extreme}

\end{document}